# Quantum Field Theory of Bose-Einstein Correlations

Leonid V. Razumov[*], R.M. Weiner [§]

Physics Department, University of Marburg, Marburg, F.R. Germany

## Abstract

A Quantum Field Theory formulation of Bose-Einstein Correlations is given. It contains as a special case the classical current approach. It is shown that the particle-antiparticle correlations are a general feature of Bose-Einstein Correlations and not an artifact of a certain approximation. Relations are provided through which the quantum corrections to the classical current approach can be determined experimentally.

[*]E.Mail: razumov@convex.hrz.Uni-Marburg.de
[§]E. Mail: weiner@vax.hrz.Uni-Marburg.de

Bose-Einstein correlations (BEC) are an important theoretical and experimental tool in particle and nuclear physics in a wide range of energies because one can obtain from BEC information about the space-time structure of the source as well as about the dynamics of the reaction mechanism.

There have been two main theoretical approaches to BEC:

1. The wave function approach [1]. This approach being based on first quantisation has serious drawbacks some of which are listed below:

   (a) The treatment of inclusive reactions which constitute the bulk of experimentally accessible reactions is quite cumbersome, if not practically impossible for high multiplicities.

   (b) Higher order correlations can hardly be calculated with this method.

   (c) A correct treatment of quantum statistical coherence is not possible.

   (d) Quantum statistical particle-antiparticle correlations are absent.

2. The classical current formalism [1, 2], which has none of the drawbacks mentioned above but still uses the approximation that the currents (but not the fields) are classical. The Wigner function approach [3] and the method of local equilibrium statistical operator [4] are particular cases of this current ensemble approach.

Recently using a particular model [5] for heavy-ion reactions it has been conjectured that measurements of a classical radius through BEC are strongly affected by non-classical off-shell effects.

In this paper we develop a quantum-field approach to BEC in the real time formulation (Keldysh technique) [6]. Among the results obtained here we mention an estimate accessible to experimental measurements of the quantum corrections to the classical current formalism and the proof that quantum-statistical particle-antiparticle correlations are not an artifact of the classical current formalism but a quite general property of particle physics.

We start with the inclusive spectra which are usually defined in terms of the density matrix $\rho_f$ of the final state

$$P_1(\bm{k}) \equiv k^0 \frac{d\sigma^{in}}{d^3\bm{k}} = T_r \rho_f a^\dagger a = \sum_f w_f n_f(\bm{k}) = <n_f(\bm{k})> \tag{1}$$

$$P_2(\bm{k}_1, \bm{k}_2) \equiv k_1^0 k_2^0 \frac{d^2\sigma^{in}}{d^3\bm{k}_1 d^3\bm{k}_2} = Tr\{\rho_f a^\dagger(\bm{k}_2) a^\dagger(\bm{k}_1) a(\bm{k}_1) a(\bm{k}_2)\} = \tag{2}$$

$$= \begin{cases} (\bm{k}_1 \neq \bm{k}_2) & ; \quad \sum_f w_f n_f(\bm{k}_1) n_f(\bm{k}_2) = <n(\bm{k}_1) n(\bm{k}_2)> \\ (\bm{k}_1 = \bm{k}_2 = \bm{k}) & ; \quad \sum_f w_f n_f(\bm{k}) \cdot (n_f(\bm{k}) - 1) = <n(\bm{k}) \cdot (n(\bm{k}) - 1)> \end{cases}$$

(where $w_f$ denotes the probability to obtain the final state $|f>$ from the initial state and $n$ is the multiplicity) and express them in terms of the initial state density matrix $\rho_i$ and the S-matrix using the identity $\rho_f = S\rho_i S^\dagger$

$$P_1(\boldsymbol{k}) = Tr\rho_i S^\dagger a^\dagger(\boldsymbol{k}) a(\boldsymbol{k}) S \tag{3}$$

$$P_2(\boldsymbol{k}_1, \boldsymbol{k}_2) \equiv k_1^0 k_2^0 \frac{d^2\sigma^{in}}{d^3\boldsymbol{k}_1 d^3\boldsymbol{k}_2} = Tr\{\rho_i S^\dagger a^\dagger(\boldsymbol{k}_2) a^\dagger(\boldsymbol{k}_1) a(\boldsymbol{k}_1) a(\boldsymbol{k}_2) S\} \tag{4}$$

To proceed further we assume [7] that in our interferometry experiment we registrate particles with quantum numbers not contained in the initial state

$$a\rho_i = \rho_i a^\dagger = 0 \tag{5}$$

This condition is in general satisfied if one omits in the final state particles with zero transverse momentum. We can transpose the annihilation operators $a$ on the r.h.s. and the creation operators $a^\dagger$ on the l.h.s. side of $Tr$ expression commuting repeatedly $a$ with $S$, $a^\dagger$ with $S^\dagger$ and using (5). This results in the following expressions for the single

$$P_1(\boldsymbol{k}) = Tr\left\{\rho_i \frac{\delta S^\dagger}{\delta a(\boldsymbol{k})} \frac{\delta S}{\delta a^\dagger(\boldsymbol{k})}\right\} \tag{6}$$

and the double inclusive cross-section

$$P_2(\boldsymbol{k}_1, \boldsymbol{k}_2) = Tr\left\{\rho_i \frac{\delta^2 S^\dagger}{\delta a(\boldsymbol{k}_1)\delta a(\boldsymbol{k}_2)} \cdot \frac{\delta^2 S}{\delta a^\dagger(\boldsymbol{k}_1)\delta a^\dagger(\boldsymbol{k}_2)}\right\} \tag{7}$$

In (6) and (7) we used the property $[a(k), S] = \delta S/\delta a^\dagger(k)$, where the variational derivative is taken with respect to the corresponding symbol of the operator $a^\dagger$ (cf. [8]).

To be specific we consider in the following the direct production of pions $\pi^{(+)}$, $\pi^{(-)}$, $\pi^0$ (the method can be easily generalized for pions coming from resonances).
We assume that the particle production is governed by the interaction lagrangian:

$$L^{int}(x) \equiv J^{(+)}(x)\pi^{(-)}(x) + J^{(-)}(x)\pi^{(+)}(x) + J^0(x)\pi^0(x) \tag{8}$$

which is the simplest ansatz for an isospin triplet. The currents $J^{(+)}$, $J^{(-)}$, $J^0$ are *operators* not depending on the $\pi^{(\pm)}$ and $\pi^0$ fields ($[J, \pi] = 0$). For example $J$ can be due to quark-fields ($J^{(-)}(x) = \bar{u}\gamma_5 d$, $J^0 = (\bar{u}\gamma_5 u - \bar{d}\gamma_5 d)/\sqrt{2}$) or any other fields different from the pion fields. In this way we generalise the classical current formalism.
Let $a(a^\dagger)$, $b(b^\dagger)$, $c(c^\dagger)$ stand for annihilation (creation) operators of $\pi^{(-)}$, $\pi^{(+)}$, $\pi^{(0)}$ respectively. The same procedure that was used to derive (6), (7) leads now to following expressions for inclusive spectra:

$$P_1^{(-)}(\boldsymbol{k}) \equiv k^0 \frac{d\sigma_{(-)}^{in}}{d^3\boldsymbol{k}} = Tr\left\{\rho_i \frac{\delta S^\dagger}{\delta a(\boldsymbol{k})} \frac{\delta S}{\delta a^\dagger(\boldsymbol{k})}\right\} \tag{9}$$

$$P_1^{(+)}(\boldsymbol{k}) \equiv k^0 \frac{d\sigma_{(+)}^{in}}{d^3\boldsymbol{k}} = Tr\left\{\rho_i \frac{\delta S^\dagger}{\delta b(\boldsymbol{k})} \frac{\delta S}{\delta b^\dagger(\boldsymbol{k})}\right\} \tag{10}$$

$$P_1^{(0)}(\boldsymbol{k}) \equiv k^0 \frac{d\sigma_{(0)}^{in}}{d^3\boldsymbol{k}} = Tr\left\{\rho_i \frac{\delta S^\dagger}{\delta c(\boldsymbol{k})} \frac{\delta S}{\delta c^\dagger(\boldsymbol{k})}\right\} \tag{11}$$

$$P_2^{(--)}(\boldsymbol{k}_1, \boldsymbol{k}_2) \equiv k_1^0 k_2^0 \frac{d^2\sigma_{(--)}^{in}}{d^3\boldsymbol{k}_1 d^3\boldsymbol{k}_2} = Tr\left\{\rho_i \frac{\delta^2 S^\dagger}{\delta a(\boldsymbol{k}_1)\delta a(\boldsymbol{k}_2)} \cdot \frac{\delta^2 S}{\delta a^\dagger(\boldsymbol{k}_1)\delta a^\dagger(\boldsymbol{k}_2)}\right\} \tag{12}$$

$$P_2^{(00)}(\boldsymbol{k}_1, \boldsymbol{k}_2) \equiv k_1^0 k_2^0 \frac{d^2\sigma_{(00)}^{in}}{d^3\boldsymbol{k}_1 d^3\boldsymbol{k}_2} = Tr\left\{\rho_i \frac{\delta^2 S^\dagger}{\delta c(\boldsymbol{k}_1)\delta c(\boldsymbol{k}_2)} \cdot \frac{\delta^2 S}{\delta c^\dagger(\boldsymbol{k}_1)\delta c^\dagger(\boldsymbol{k}_2)}\right\} \tag{13}$$

$$P_2^{(-+)}(\boldsymbol{k}_1, \boldsymbol{k}_2) \equiv k_1^0 k_2^0 \frac{d^2\sigma_{(-+)}^{in}}{d^3\boldsymbol{k}_1 d^3\boldsymbol{k}_2} = Tr\left\{\rho_i \frac{\delta^2 S^\dagger}{\delta a(\boldsymbol{k}_1)\delta b(\boldsymbol{k}_2)} \cdot \frac{\delta^2 S}{\delta a^\dagger(\boldsymbol{k}_1)\delta b^\dagger(\boldsymbol{k}_2)}\right\} \tag{14}$$

Formulae (9)-(11) represent single inclusive cross-sections of $\pi^{(-)}$, $\pi^{(+)}$ and $\pi^0$. Double inclusive spectra for $\pi^-\pi^- = \pi^{(+)}\pi^{(+)}$ and $\pi^{(0)}\pi^{(0)}$ pairs are described by (12) and (13) respectively. Formula (14) stands for the $\pi^{(-)}\pi^{(+)}$ double inclusive cross-section, which reflects Bose-Einstein Correlations between a particle and its antiparticle [2].

Next we eliminate the field operators in favour of the current operators $J$ by means of the $S$-matrix represented through the interaction lagrangian

$$S = \mathcal{T}\exp\left\{i\int d^4x L^{int}(x)\right\} \tag{15}$$

and calculate the derivatives:

$$\frac{\delta S}{\delta a^\dagger(\boldsymbol{k})} = \frac{\mathcal{T}\{J^{(-)}(k)S\}}{\sqrt{2k^0 V}}, \qquad \frac{\delta S^\dagger}{\delta a(\boldsymbol{k})} = \frac{\tilde{\mathcal{T}}\{S^\dagger J^{(+)}(-k)\}}{\sqrt{2k^0 V}},$$

$$\frac{\delta S}{\delta b^\dagger(\boldsymbol{k})} = \frac{\mathcal{T}\{J^{(+)}(k)S\}}{\sqrt{2k^0 V}}, \qquad \frac{\delta S^\dagger}{\delta b(\boldsymbol{k})} = \frac{\tilde{\mathcal{T}}\{S^\dagger J^{(-)}(-k)\}}{\sqrt{2k^0 V}}, \tag{16}$$

$$\frac{\delta S}{\delta c^\dagger(\boldsymbol{k})} = \frac{\mathcal{T}\{J^{(0)}(k)S\}}{\sqrt{2k^0 V}}, \qquad \frac{\delta S^\dagger}{\delta c(\boldsymbol{k})} = \frac{\tilde{\mathcal{T}}\{S^\dagger J^{(0)}(-k)\}}{\sqrt{2k^0 V}},$$

Here T and $\tilde{\mathcal{T}}$ are the chronological and antichronological time ordering operators respectively. No difficulties arise for the second and higher order derivatives because the currents $J$ do not depend on $a$, $a^\dagger$, $b$, $b^\dagger$, $c$, $c^\dagger$.

The single and double inclusive spectra read now:

$$P_1^{(-)}(\boldsymbol{k}) = Tr\left\{\rho_i \tilde{\mathcal{T}}\left[S^\dagger J^{(+)}(-k)\right]\mathcal{T}\left[J^{(-)}(k)S\right]\right\} \tag{17}$$

$$P_1^{(+)}(\boldsymbol{k}) = Tr\left\{\rho_i \tilde{\mathcal{T}}\left[S^\dagger J^{(-)}(-k)\right] \mathcal{T}\left[J^{(+)}(k) S\right]\right\} \tag{18}$$

$$P_1^{(0)}(\boldsymbol{k}) = Tr\left\{\rho_i \tilde{\mathcal{T}}\left[S^\dagger J^{(0)}(-k)\right] \mathcal{T}\left[J^{(0)}(k) S\right]\right\} \tag{19}$$

$$P_2^{(--)}(\boldsymbol{k}_1, \boldsymbol{k}_2) = Tr\left\{\rho_i \tilde{\mathcal{T}}\left[S^\dagger J^{(+)}(-k_1) J^{(+)}(-k_2)\right] \mathcal{T}\left[J^{(-)}(k_1) J^{(-)}(k_2) S\right]\right\} \tag{20}$$

$$P_2^{(00)}(\boldsymbol{k}_1, \boldsymbol{k}_2) = Tr\left\{\rho_i \tilde{\mathcal{T}}\left[S^\dagger J^{(0)}(-k_1) J^{(0)}(-k_2)\right] \mathcal{T}\left[J^{(0)}(k_1) J^{(0)}(k_2) S\right]\right\} \tag{21}$$

$$P_2^{(-+)}(\boldsymbol{k}_1, \boldsymbol{k}_2) = Tr\left\{\rho_i \tilde{\mathcal{T}}\left[S^\dagger J^{(+)}(-k_1) J^{(-)}(-k_2)\right] \mathcal{T}\left[J^{(-)}(k_1) J^{(+)}(k_2) S\right]\right\} \tag{22}$$

We have thus obtained the expressions for the physical quantities only in terms of the currents and $S$-matrix. To eliminate the $S$-matrix we rewrite (17) - (22) in terms of *Heisenberg* operators. To do so we insert the identity $\mathbf{1} = S S^\dagger$ between the antichronological $\tilde{\mathcal{T}}$ and chronological $\mathcal{T}$ parts of expressions (17) - (22) and use the well known relation between *Heisenberg* and *Interaction* representation operators:

$$\mathcal{T}\left[A_H B_H C_H \ldots\right] = S^\dagger \mathcal{T}\left[A_{int} B_{int} C_{int} \ldots S\right]$$

and $\tag{23}$

$$\tilde{\mathcal{T}}\left[A_H B_H C_H \ldots\right] = \mathcal{T}\left[S^\dagger A_{int} B_{int} C_{int} \ldots\right] S$$

This procedure results in the following formulae:

$$P_1^{(-)}(\boldsymbol{k}) = Tr\left\{\rho_i J_H^{(+)}(-k) J_H^{(-)}(k)\right\} \tag{24}$$

$$P_1^{(+)}(\boldsymbol{k}) = Tr\left\{\rho_i J_H^{(-)}(-k) J_H^{(+)}(k)\right\} \tag{25}$$

$$P_1^{(0)}(\boldsymbol{k}) = Tr\left\{\rho_i J_H^{(0)}(-k) J_H^{(0)}(k)\right\} \tag{26}$$

$$P_2^{(--)}(\boldsymbol{k}_1, \boldsymbol{k}_2) = Tr\left\{\rho_i \tilde{\mathcal{T}}\left[J_H^{(+)}(-k_1) J_H^{(+)}(-k_2)\right] \mathcal{T}\left[J_H^{(-)}(k_1) J_H^{(-)}(k_2)\right]\right\} \tag{27}$$

$$P_2^{(00)}(\boldsymbol{k}_1, \boldsymbol{k}_2) = Tr\left\{\rho_i \tilde{\mathcal{T}}\left[J_H^{(0)}(-k_1) J_H^{(0)}(-k_2)\right] \mathcal{T}\left[J_H^{(0)}(k_1) J_H^{(0)}(k_2)\right]\right\} \tag{28}$$

$$P_2^{(-+)}(\boldsymbol{k}_1, \boldsymbol{k}_2) = Tr\left\{\rho_i \tilde{\mathcal{T}}\left[J_H^{(+)}(-k_1) J_H^{(-)}(-k_2)\right] \mathcal{T}\left[J_H^{(-)}(k_1) J_H^{(+)}(k_2)\right]\right\} \tag{29}$$

Similar expressions can be derived of course for higher order inclusive cross-sections. From now on we shall omit the index $H$ assuming that all operators are in the Heisenberg representation. Eqs. (24)-(29) constitute the Quantum Field theoretical formulation of the inclusive spectra in terms of quantum currents and the density matrix of the initial state.

For the particular case of recoilless emission the quantum currents $J$ can be replaced by classical currents [9] which simplifies to a great extent the expressions above. Thus e.g. for $\pi^{(+)}\pi^{(-)}$ spectra one obtains as in [2]

$$P_{2_{\text{cl}}}^{(-+)}(\boldsymbol{k}_1, \boldsymbol{k}_2) = \int DJ^{(-)} DJ^{(+)} \mathsf{P}(J^{(-)}, J^{(+)}) \times \tag{30}$$

$$\times <J^{(+)}(-\boldsymbol{k}_1) J^{(-)}(-\boldsymbol{k}_2) J^{(-)}(\boldsymbol{k}_1) J^{(+)}(\boldsymbol{k}_2)>$$

where the trace over the density matrix $\rho_i$ has been replaced by an integral over the classical current configurations.

In Quantum Field Theory there is a very elegant formalism to deal with antichronologically ordered operator products on the same footing as with chronological ones, namely the Keldysh technique [6] well known also under the name Real Time Formalism (RTF) The essence of this formalism consists in the following:

Every dynamical operator $A$ is mapped formally in two operators denoted here as $A_\oplus$, $A_\ominus$. These additional degrees of freedom reflect what kind of ordering is used (Ex.: $\mathcal{T}_c[A_\ominus B_\ominus] = \mathcal{T}[AB]$; $\mathcal{T}_c[A_\oplus B_\oplus] = \tilde{\mathcal{T}}[AB]$; $\mathcal{T}_c[A_\oplus B_\ominus] = AB$; $\mathcal{T}_c[A_\ominus B_\oplus] = BA$). In RTF the double-inclusive spectra (27)-(29) are just "*contour-time*" ($\mathcal{T}_c$) ordered Green-function of the corresponding operators:

$$P_2^{(--)}(\boldsymbol{k}_1, \boldsymbol{k}_2) = <\mathcal{T}_c \left\{ J_\oplus^{(+)}(-k_1) J_\oplus^{(+)}(-k_2) J_\ominus^{(-)}(k_1) J_\ominus^{(-)}(k_2) \right\}> \tag{31}$$

$$P_2^{(00)}(\boldsymbol{k}_1, \boldsymbol{k}_2) = <\mathcal{T}_c \left\{ J_\oplus^{(0)}(-k_1) J_\oplus^{(0)}(-k_2) J_\ominus^{(0)}(k_1) J_\ominus^{(0)}(k_2) \right\}> \tag{32}$$

$$P_2^{(-+)}(\boldsymbol{k}_1, \boldsymbol{k}_2) = <\mathcal{T}_c \left\{ J_\oplus^{(+)}(-k_1) J_\oplus^{(-)}(-k_2) J_\ominus^{(-)}(k_1) J_\ominus^{(+)}(k_2) \right\}> \tag{33}$$

Above we have introduced the notation $<\ldots> \equiv Tr\{\rho_i \ldots\}$. Formally these equations are quite similar to those obtained in the classical current formalism (compare e.g. Eq. (33) with Eq. (30)); the difference resides, of course, in the fact that in the quantum-field formulation the currents are operators and they appear in a particular order.

It is useful to rewrite the full Green-functions (denoted by $<\ldots>$) in terms of their connected parts (denoted here as $\ll \ldots \gg$). For simplicity we assume from now on that the system is completely chaotic ($<J> = 0$). Then, the double-inclusive spectra (31)-(33) written in terms of *connected* Green-functions take the following form (for their graphic representation (cf. Fig.1):

$$P_2^{(--)}(\boldsymbol{k}_1, \boldsymbol{k}_2) = \ll \mathcal{T}_c \left\{ J_\oplus^{(+)}(-k_1) J_\oplus^{(+)}(-k_2) J_\ominus^{(-)}(k_1) J_\ominus^{(-)}(k_2) \right\} \gg +$$

$$+ \ll \mathcal{T}_c \left\{ J_\oplus^{(+)}(-k_1) J_\ominus^{(-)}(k_1) \right\} \gg \ll \mathcal{T}_c \left\{ J_\oplus^{(+)}(-k_2) J_\ominus^{(-)}(k_2) \right\} \gg$$

$$+ \ll \mathcal{T}_c \left\{ J_\oplus^{(+)}(-k_1) J_\ominus^{(-)}(k_2) \right\} \gg \ll \mathcal{T}_c \left\{ J_\oplus^{(+)}(-k_2) J_\ominus^{(-)}(k_1) \right\} \gg \tag{34}$$

$$P_2^{(00)}(\boldsymbol{k}_1, \boldsymbol{k}_2) = \ll \mathcal{T}_c \left\{ J_\oplus^{(0)}(-k_1) J_\oplus^{(0)}(-k_2) J_\ominus^{(0)}(k_1) J_\ominus^{(0)}(k_2) \right\} \gg$$

$$+ \ll \mathcal{T}_c \left\{ J_\oplus^{(0)}(-k_1) J_\ominus^{(0)}(k_1) \right\} \gg \ll \mathcal{T}_c \left\{ J_\oplus^{(0)}(-k_2) J_\ominus^{(0)}(k_2) \right\} \gg$$

$$+ \ll \mathcal{T}_c \left\{ J_\oplus^{(0)}(-k_1) J_\ominus^{(0)}(k_2) \right\} \gg \ll \mathcal{T}_c \left\{ J_\oplus^{(0)}(-k_2) J_\ominus^{(0)}(k_1) \right\} \gg$$

$$+ \ll \mathcal{T}_c \left\{ J_\oplus^{(0)}(-k_1) J_\oplus^{(0)}(-k_2) \right\} \gg \ll \mathcal{T}_c \left\{ J_\ominus^{(0)}(k_2) J_\ominus^{(0)}(k_1) \right\} \gg \quad (35)$$

$$P_2^{(-+)}(\boldsymbol{k}_1, \boldsymbol{k}_2) = \ll \mathcal{T}_c \left\{ J_\oplus^{(+)}(-k_1) J_\oplus^{(-)}(-k_2) J_\ominus^{(-)}(k_1) J_\ominus^{(+)}(k_2) \right\} \gg$$

$$+ \ll \mathcal{T}_c \left\{ J_\oplus^{(+)}(-k_1) J_\ominus^{(-)}(k_1) \right\} \gg \ll \mathcal{T}_c \left\{ J_\oplus^{(-)}(-k_2) J_\ominus^{(+)}(k_2) \right\} \gg$$

$$+ \ll \mathcal{T}_c \left\{ J_\oplus^{(+)}(-k_1) J_\oplus^{(-)}(-k_2) \right\} \gg \ll \mathcal{T}_c \left\{ J_\ominus^{(-)}(k_1) J_\ominus^{(+)}(k_2) \right\} \gg \quad (36)$$

The expressions (34)-(36) are exact but quite complicated.

An important simplification can be achieved by assuming a Gaussian form for the density matrix [10]. This form follows from the Central Limit Theorem which is applicable when the number of independent sources becomes large. The assumption that the number of sources is large is quite natural for high energy heavy-ion reactions and it is an accepted ingredient in many phenomenological approaches like e.g. the string model [11, 12], hydrodynamical [13] and thermodynamical models, etc. For a gaussian density matrix the 4-point connected correlator vanishes. Then eqs. (24) - (26), (34)-(36) become

$$P_1^{(0)}(\boldsymbol{k}) = F^n(k, k) \quad (37)$$

$$P_1^{(-)}(\boldsymbol{k}) = P_1^{(+)}(\boldsymbol{k}) = F^{ch}(k, k) \quad (38)$$

$$P_2^{(--)}(\boldsymbol{k}_1, \boldsymbol{k}_2) = P_1^{(-)}(\boldsymbol{k}_1) P_1^{(-)}(\boldsymbol{k}_2) + |F^{ch}(k_1, k_2)|^2 \quad (39)$$

$$P_2^{(00)}(\boldsymbol{k}_1, \boldsymbol{k}_2) = P_1^{(0)}(\boldsymbol{k}_1) P_1^{(0)}(\boldsymbol{k}_2) + |F^n(k_1, k_2)|^2 + |\tilde{F}^n(k_1, k_2)|^2 \quad (40)$$

$$P_2^{(-+)}(\boldsymbol{k}_1, \boldsymbol{k}_2) = P_1^{(-)}(\boldsymbol{k}_1) P_1^{(+)}(\boldsymbol{k}_2) + |\tilde{F}^{ch}(-k_1, k_2)|^2 \quad (41)$$

where the functions $F$ are defined for charged particles (upper index $ch$) and for neutral ones (upper index $n$) as follows:

$$F^{ch}(k_1, k_2) \equiv \ll \mathcal{T}_c \left\{ J_\oplus^{(+)}(-k_1) J_\ominus^{(-)}(k_2) \right\} \gg = \ll J^{(+)}(-k_1) J^{(-)}(k_2) \gg \quad (42)$$

$$\tilde{F}^{ch}(k_1, k_2) \equiv \ll \mathcal{T}_c \left\{ J_\oplus^{(+)}(-k_1) J_\oplus^{(-)}(k_2) \right\} \gg = \ll \tilde{\mathcal{T}} \left\{ J^{(+)}(-k_1) J^{(-)}(k_2) \right\} \gg \quad (43)$$

$$F^n(k_1, k_2) \equiv \ll \mathcal{T}_c \left\{ J_\oplus^{(0)}(-k_1) J_\ominus^{(0)}(k_2) \right\} \gg = \ll J^{(0)}(-k_1) J^{(0)}(k_2) \gg \quad (44)$$

$$\tilde{F}^n(k_1, k_2) \equiv \ll \mathcal{T}_c \left\{ J_\oplus^{(0)}(-k_1) J_\oplus^{(0)}(k_2) \right\} \gg \, = \, \ll \tilde{\mathcal{T}} \left\{ J^{(0)}(-k_1) J^{(0)}(k_2) \right\} \gg \quad (45)$$

In contrast to the classical current approach [2] which deals with only one type of two-current correlator we have here two different kinds of two-current correlators depending on their ordering prescriptions. The two-particle correlations are given by

$$C_2^{(-+)}(\boldsymbol{k}_1, \boldsymbol{k}_2) \equiv \frac{P_2^{(-+)}(\boldsymbol{k}_1, \boldsymbol{k}_2)}{P_1^{(-)}(\boldsymbol{k}_1) P_1^{(+)}(\boldsymbol{k}_2)} = 1 + |\tilde{d}^{ch}(k_1, k_2)|^2 \quad (46)$$

$$C_2^{(--)}(\boldsymbol{k}_1, \boldsymbol{k}_2) \equiv \frac{P_2^{(--)}(\boldsymbol{k}_1, \boldsymbol{k}_2)}{P_1^{(-)}(\boldsymbol{k}_1) P_1^{(-)}(\boldsymbol{k}_2)} = 1 + |d^{ch}(k_1, k_2)|^2 \quad (47)$$

$$C_2^{(00)}(\boldsymbol{k}_1, \boldsymbol{k}_2) \equiv \frac{P_2^{(00)}(\boldsymbol{k}_1, \boldsymbol{k}_2)}{P_1^0(\boldsymbol{k}_1) P_1^0(\boldsymbol{k}_2)} = 1 + |d^n(k_1, k_2)|^2 + |\tilde{d}^n(k_1, k_2)|^2 \quad (48)$$

where the functions $d$ are defined as follows:

$$d^{ch}(k_1, k_2) \equiv \frac{F^{ch}(k_1, k_2)}{\sqrt{F^{ch}(k_1, k_1) F^{ch}(k_2, k_2)}} \quad ; \quad \tilde{d}^{ch}(k_1, k_2) \equiv \frac{\tilde{F}^{ch}(k_1, k_2)}{\sqrt{F^{ch}(k_1, k_1) F^{ch}(k_2, k_2)}} \quad (49)$$

$$d^n(k_1, k_2) \equiv \frac{F^n(k_1, k_2)}{\sqrt{F^n(k_1, k_1) F^n(k_2, k_2)}} \quad ; \quad \tilde{d}^n(k_1, k_2) \equiv \frac{\tilde{F}^n(k_1, k_2)}{\sqrt{F^n(k_1, k_1) F^n(k_2, k_2)}} \quad (50)$$

The "surprising" effects found in [2] i.e. the

- presence of particle-antiparticle Bose-Einstein type correlations and

- a new term in Bose-Einstein correlation function for neutral particles

are reobtained, but under a more general form which contains also the quantum corrections. These equations also prove that the effects derived in [2] are not an artifact of the classical current formalism but have general validity.

To study the deviations of the above equations from the classical current approach we compare the two types of two-current correlators (42) and (43). This can be done more easily in coordinate representation ($F^{ch}(x_1, x_2) = \ll J_\oplus^{(+)}(x_1) J_\ominus^{(-)}(x_2) \gg ; \tilde{F}^{ch}(x_1, x_2) = \ll \tilde{\mathcal{T}} \{ J_\oplus^{(+)}(x_1) J_\oplus^{(-)}(x_2) \} \gg$):

$$R^{ch}(x_1, x_2) \equiv \tilde{F}^{ch}(x_1, x_2) - F^{ch}(x_1, x_2) = \Theta(x_1^0 - x_2^0) \ll [J^{(+)}(x_1) J^{(-)}(x_2)] \gg \quad (51)$$

The $R^{ch}$-function is just the retarded current commutator and can in principle be calculated once the currents and fields entering (8) are given (cf. the end of the paper). In principle, one can determine $R^{ch}$ experimentally by comparing $\pi^{(-)} \pi^{(-)}$ and $\pi^{(-)} \pi^{(+)}$ data and using Eqs. (38), (46), (47), (49), (51). In this way one can test to what extent the system can be treated quasi-classically. On the other hand the $R$-function can be calculated for a particular model. For any classical current theory $R = 0$ and therefore $\tilde{d}(k_1, k_2) = d(k_1, -k_2)$ [14].

Similarly for the neutral particles one has

$$R^n(x_1, x_2) \equiv \tilde{F}^n(x_1, x_2) - F^n(x_1, x_2) = \Theta(x_1^0 - x_2^0) \ll [J^{(0)}(x_1), J^{(0)}(x_2)] \gg \tag{52}$$

which is the retarded commutator of the current $J^{(0)}$.

The structure of formula (48) is typical for the correlation function of any neutral boson. For instance, if one measures photon-pairs with a given polarisation, formula (48) is applicable if one defines $J^{(0)}(k) \equiv \epsilon^\mu(k) J^\mu(k)$. In general polarisation effects for particles with non-zero integer spin can play an important role in (40) [15]. But apart of polarisation effects the kinematical and dynamical information about the collision is contained in the two functions $F^n$ and $\tilde{F}^n$.

Since $|d^n(k,k)| = 1$ and $|\tilde{d}^n(k_1, k_2)| \geq 0$ the intercept of the correlation function $C_2^{(00)}(k,k) \geq 2$ (we refer here to a completely chaotic source, without coherence). The exact value of the intercept is model dependent. As mentioned above, in the classical current formalism $R \equiv 0$, $\tilde{d}^n(k_1, k_2) = d^n(-k_1, k_2)$ and the correlation function simplifies to

$$C_{2\text{cl}}^{(00)}(\mathbf{k}_1, \mathbf{k}_2) = 1 + |d^n(k_1, k_2)|^2 + |d^n(-k_1, k_2)|^2 \tag{53}$$

which depends only on the function $d^n$. For this case the bounds for the correlation function have been established [16][1].

To exemplify the above made considerations we present here a simple model where massless pions are produced from a static source consisting of other scalar particles in thermal equilibrium ($\rho = e^{-H/T}$ where $H$ is the Hamiltonian of the system and $T$ is the temperature, cf. e.g. [18]). The interaction is given by the Lagrangian

$$L_{int} = g j(x) \pi(x) \quad , \quad \text{with} \quad j(x) = \bar{\varphi}(x) \varphi(x) \tag{54}$$

The calculations are performed in the one-loop approximation using the Keldysh technique [6].

The relevant current correlators can be expressed in a very simple form if one takes

---

[1] In [17] the results of the classical current formalism were misquoted. Contrary to what was stated in [17] the intercept of the $\pi^0 \pi^0$ correlation function in the classical current formalism strongly depends on both the mass $m$ of the particle and the time-duration of freezeout $\tau$. The value 3 for the intercept can be obtained either in the limit of instantaneous freezeout or for massless particles at zero momentum. For the "static" source with exponential parametrisation one has $C_2^{\max} = 2 + \exp(-\tau m)$.

into account the contour time ordering by placing $(+-)$ labels at the vertices:

$$< j(x_1)j(x_2) > = \quad \underset{x_1}{(+)} \bullet \!\!\!\!\!\!\!\!\!\!\!\!\!\!\!\!\!\!\!\!\!\!\!\!\!\!\!\!\!\!\!\!\!\!\!\!\!\!\!\!\!\!\!\!\!\!\!\!\!\!\!\!\! \underset{x_2}{(-)} \bullet \quad = \quad (< \varphi(x_1)\bar{\varphi}(x_2) >)^2$$

$$< j(x_2)j(x_1) > = \quad \underset{x_1}{(-)} \bullet \!\!\!\!\!\!\!\!\!\!\!\!\!\!\!\!\!\!\!\!\!\!\!\!\!\!\!\!\!\!\!\!\!\!\!\!\!\!\!\!\!\!\!\!\!\!\!\!\!\!\!\!\! \underset{x_2}{(+)} \bullet \quad = \quad (< \varphi(x_2)\bar{\varphi}(x_1) >)^2$$

where
$$< \varphi(x_1)\bar{\varphi}(x_2) > = \frac{1}{\hbar} \int \frac{d^4p}{(2\pi)^4} e^{-ip(x_1-x_2)} 2\pi \delta(p^2)[\Theta(p_0) + n(|p_0|)] \tag{55}$$

The $R$-function which is the retarded current commutator (52) can be calculated even easier than the correlators

$$[j(x_1), j(x_2)] = [\bar{\varphi}(x_1)\varphi(x_1) , \bar{\varphi}(x_2)\varphi(x_2)] \tag{56}$$

$$= \bar{\varphi}(x_1)[\varphi(x_1) , \bar{\varphi}(x_2)]\varphi(x_2) + \bar{\varphi}(x_2)[\bar{\varphi}(x_1) , \varphi(x_2)]\varphi(x_1)$$

because the field commutator $[\varphi(x_1) , \bar{\varphi}(x_2)]$ does not depend on temperature:

$$[\varphi(x_1) , \bar{\varphi}(x_2)] = \frac{1}{\hbar} \int \frac{d^4p}{(2\pi)^4} e^{-ip(x_1-x_2)} 2\pi \mathrm{sgn}(p_0)\delta(p^2) \tag{57}$$

We omit the algebraic details and present just the results:

$$R(x_1, x_2) = \frac{-i}{2\pi^2} g^2 \hbar T f(x_1) \Theta(x_1^0 - x_2^0) \frac{\delta(x^2)}{|ux|} f(x_2) \left[ \frac{1}{e^{2\pi a} - 1} - \frac{1}{2\pi a} + \frac{1}{2} \right] \tag{58}$$

$$< j(x_1)j(x_2) > = f(x_1)g^2 \left\{ \frac{-i}{2\pi^2} \frac{\hbar}{x^2 - i\epsilon \mathrm{sgn}(x_0)} + \right. \tag{59}$$

$$\left. \frac{T}{2\pi} \int_{-\infty}^{+\infty} d\alpha \, \mathrm{sgn}(\alpha - (ux))\delta\left((\alpha u^\mu - x^\mu)^2\right) \left[ \frac{1}{e^{2\pi\alpha T/\hbar} - 1} - \frac{1}{2\pi\alpha T/\hbar} + \frac{1}{2} \right] \right\}^2 f(x_2)$$

where $x \equiv x_1 - x_2$ and we define $a \equiv 2T|ux|/\hbar$ using the 4-velocity $u^\mu$ ( in rest frame $u^\mu = (1, \mathbf{0})$). The function $f(x)$ is introduced phenomenologically and plays the role of an effective vertex function containing the finite size effects of the source. It is the subject of Bose-Einstein correlations studies.

It is easy to see from (58) and (59) that this model manifests quasi-classical behaviour (that is $|R(x_1, x_2)/<j(x_1)j(x_2)>| \ll 1$) at least in two limiting cases: $T \to \infty$ or $\hbar \to 0$ as it should be from general principles.

Although this model can not be considered as realistic it illustrates how quantum effects do influence BEC.

Finally one should stress a new feature of particle-antiparticle BEC which emerges from this paper. Besides the fact that like ordinary particle-particle BEC they can serve as tools in the determination of radii and lifetimes, and besides the fact that they reflect the existence of squeezed states, they play a particularly important role in the detection of quantum corrections to BEC. Their experimental measurement is a highly rewarding task.

We would like to acknowledge fruitful comments by I.V.Andreev, C.Jeffries, H.Paul, M.Plümer. This work was supported in part by the Gesellschaft für Schwerionenforschung, Darmstadt.

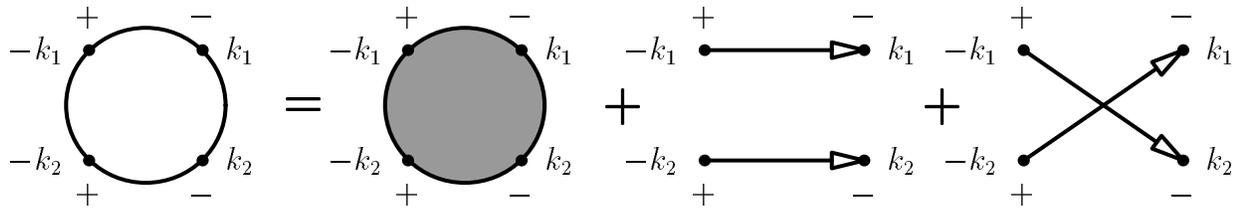

**Fig. 1**  Decomposition of the $\pi^{(-)}\pi^{(-)}$ 4-point full Green-function.
in terms of the connected parts. This diagram corresponds to $P_2^{(--)}(k_1, k_2)$.

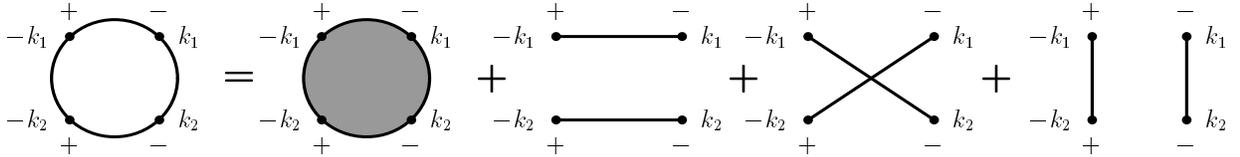

**Fig. 2**  Decomposition of the $\pi^{(0)}\pi^{(0)}$ 4-point full Green-function
in terms of the connected parts. This diagram corresponds to $P_2^{(00)}(k_1, k_2)$.

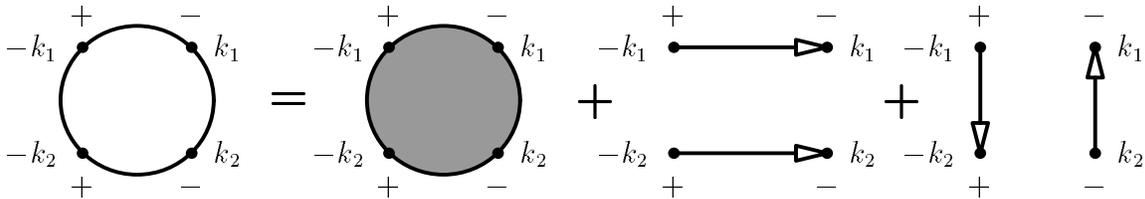

**Fig. 3**  Decomposition of the $\pi^{(-)}\pi^{(+)}$ 4-point full Green-function
in terms of the connected parts. This diagram corresponds to $P_2^{(-+)}(k_1, k_2)$.